**Title**

Emulating Stepped-Wedge Cluster Randomized Trials to Evaluate Health Policies and Interventions

**Short Title**

Emulating Stepped-Wedge Cluster Randomized Trials

**Authors**

Haidong Lu[1,2,3,4]; Gregg S. Gonsalves[3,4,5]; Fan Li[6,7]; Guangyu Tong[1,6,7,8]; Lee Kennedy-Shaffer[6]*

[1] Department of Internal Medicine, Yale School of Medicine, New Haven, CT, US

[2] Program in Addiction Medicine, Yale School of Medicine, New Haven, CT, US

[3] Public Health Modeling Unit, Yale School of Public Health, New Haven, CT, US

[4] Department of Chronic Disease Epidemiology, Yale School of Public Health, New Haven, CT, US

[5] Department of Epidemiology of Microbial Diseases, Yale School of Public Health, New Haven, CT, US

[6] Department of Biostatistics, Yale School of Public Health, New Haven, CT, US

[7] Center for Methods in Implementation and Prevention Science, Yale School of Public Health, New Haven, CT, US

[8] Cardiovascular Medicine Analytics Center, Yale School of Medicine, New Haven, CT, US

* Corresponding author. Email: lee.kennedy-shaffer@yale.edu

**Abstract**

Both cluster randomized trials and quasi-experimental designs are used to evaluate the impact of health and social policies and interventions. Stepped-wedge cluster randomized trials randomize a staggered adoption approach, while recent difference-in-differences methods allow analysis of non-randomized settings where similar policies are adopted at different time points. These approaches have become common, but the sheer variety of methods for analyzing observational studies with staggered adoption makes it challenging to clearly design and report such studies. We propose that observational and quasi-experimental study investigators can address these challenges by emulating stepped-wedge cluster randomized trials in the target trial emulation framework. The conceptual framework and reporting standards of trial emulation will encourage consideration of key features of these designs, such as policy heterogeneity and time-varying effects, and clear reporting of the estimand and assumptions. It also highlights areas where those interested in randomized trials and quasi-experimental designs can benefit from one another's experience by bringing insights across disciplines. Questions of treatment effect heterogeneity, power, spillovers, and anticipation effects, among others, are common to both fields and can benefit from cross-pollination. This article also demonstrates how trial emulation can identify settings that are not well-served by either approach, thereby avoiding studies unlikely to generate high-quality causal evidence. Finally, it informs the bias-variance-generalizability trade-off that arises with design and analysis choices made in these settings, supporting better evidence generation and interpretation in settings where important questions can be answered.

**Introduction**

Evaluating policy interventions is an important aspect of improving health policy and ensuring effective public health responses at the institutional, municipal, national, and global levels. Policy evaluation, especially for health and social policies, is often limited by the impracticality of randomization. Therefore, observational studies are an important source of evidence on the effectiveness of such policies (1,2). The availability of large-scale population health datasets has enabled researchers to investigate these questions; however, observational studies using these data face a variety of threats to both internal and external validity that must be carefully evaluated by researchers and readers (1,3).

Quasi-experiments are observational studies that exploit variation in intervention exposure (i.e., policy adoption or implementation) across units and time points that arises independently of researcher control (2). These can be analyzed through a variety of approaches, including confounder-control and instrument-based methods (1,2). Staggered adoption designs have become increasingly common among difference-in-differences (DID) and synthetic control studies, as political units or other entities adopt similar policies at varying time points (4). Analysis methods have been adapted for these staggered settings, with adjustments made to handle biases that can arise with staggered adoption (4–9). With these new methods comes researcher choice in assumptions, methods, and reporting.

For health interventions on populations, cluster randomized trials (CRTs) enable evaluation of population- and group-level effects while accounting for within-cluster correlation (10). CRTs describe a class of designs wherein randomization occurs at the cluster (or group) level, with analysis of either individual- or cluster-level outcomes depending on the desired estimand (11). These can be parallel-arm CRTs, where clusters are randomized to intervention or control simultaneously at the start of the study, or longitudinal CRTs with repeated measurement of the outcome in clusters (although not necessarily the same individuals) over time, often with crossovers of clusters between treatment arms (10,12).

SW-CRTs are a subset of longitudinal CRTs wherein each cluster is randomized to a treatment sequence. Clusters begin the trial in the control condition and cross over to the intervention at the time point determined by their randomized sequence, with outcome measurements collected—among a closed cohort, open cohort, or repeated cross-section of individuals—at regular intervals before and after this crossover (13). SW-CRTs may also permit some clusters to never cross over to intervention during the study, sometimes referred to as hybrid SW-CRTs (14). CRTs and SW-CRTs can be particularly useful as pragmatic trials using naturally-occurring institutions or regions as clusters (10,15).

For observational studies of individual-level interventions using electronic health records and other routinely collected data, the target trial emulation (TTE) framework has become popular for clarifying and justifying the investigators' goals, assumptions, and methods, and for encouraging transparent reporting (16–19). TTE formalizes the design, analysis, and reporting of non-randomized studies by emulating key design and analysis features with reference to an idealized target trial (16,20). It requires explicit specification of key design and analysis elements for both the target trial (which may be real or hypothetical) and the observational study analog or emulation (16,19,21–23). Emphasis is also placed on the specification of "time zero" to avoid time-related biases and the specification of the estimand and population to inform appropriate interpretation and identify threats to internal and external validity (20,23–26). This further aligns understanding between those more familiar with experimental and observational approaches (27).

For health policy evaluations using DID and related observational analysis methods, Ben-Michael et al. (21) and Seewald et al. (22) recommend using TTE, specifically emulating parallel-arm CRTs when the units of implementation and analysis differ. They propose defining time zero as the time of policy adoption by the exposed unit(s) and aligning to that time point for analysis. For staggered adoption settings, this process is repeated at each adoption time point, creating a sequence of "nested" or "stacked" target trials, analogous to sequential nested trial design that is commonly employed in TTE for individual-level interventions (21,22,24,28). This design provides a useful approach that aligns with some staggered adoption DID methods (4,6).

In this article, we propose a unified conceptual framework that ties together these three threads of active research: staggered adoption quasi-experiments, SW-CRT design and analysis, and TTE. We extend the perspective of Ben-Michael et al. (21) and Seewald et al. (22) by proposing the emulation of SW-CRTs with observational panel data. Design and analysis proceed with staggered adoption on a natural time scale, rather than aligning to different time zeroes. This exploits the similarities between different staggered adoption settings noted previously (3,29).

Our approach allows researchers to articulate the research question and target estimand more clearly, transparently state identifying assumptions, choose appropriate analysis methods, and interpret the results carefully, especially in light of policy variation and time-varying effects. It focuses on maximizing information to target desired estimands by incorporating principles from the SW-CRT literature and allows extensions to other longitudinal CRT designs. We describe this approach in the next section, provide two illustrative examples in the following section, and conclude with a discussion on the value of uniting these fields to improve causal inference for social policies and health interventions.

**Emulating Stepped-Wedge Cluster Randomized Trials**

We propose an alternative TTE for staggered adoption policy evaluation to that of Ben-Michael et al. and Seewald et al. (21,22), using a SW-CRT as the target trial, rather than stacked parallel-arm CRTs. While both approaches can target similar causal estimands, when the time effects are pronounced and timing variation potentially meaningful, the SW-CRT emulation identifies important considerations for bias and power. For the components of the emulation, we generally follow those described in the TARGET Statement (19), which broadly align with other TTE implementations (17,21–23).

*Component 1. Eligibility Criteria*. For the target trial, the investigator specifies the units of randomization (i.e., clusters) and analysis (clusters or individuals).(30,31) For the observational analog, these are the policy-level units and the units on whom outcomes are measured, respectively. In addition, any eligibility criteria should be specified, including time-varying eligibility (32).

*Component 2. Treatment Strategies*. For the target trial, intervention and control policies are specified. For the observational analog, the set of policies included in each treatment category (i.e., exposed or unexposed) are specified. For heterogeneous policies, a minimum acceptable set of features is specified. This defines the consistency or stable unit treatment value assumptions (33) by articulating which cluster-periods will be considered as exposed or unexposed.

*Component 3. Assignment Procedures*. For the target trial, the type of randomization (i.e., the appropriate CRT design) is specified. The observational study will have non-random policy adoption in the specified units. Ensuring positivity—that both the exposed and unexposed conditions have some cluster-periods—is necessary if the assignment is not controlled by investigators. Understanding how the observed assignment occurred, and the implications of this for methods to achieve exchangeability, is key here.

*Component 4. Follow-Up*. The timing of the target trial, which includes the full date range of outcomes that would be included for analysis (both prior to the first intervention time point and after full adoption, if applicable) in the design, is specified. This includes whether the primary measure of time will be defined by calendar time or some other measure (e.g., epidemic time) (21). In addition, any excluded periods around intervention adoption due to anticipation, implementation, wash-out, or carry-over effects should be pre-specified (34,35). The observational analog should match these decisions as closely as possible.

In the stacked or nested parallel-arm CRT emulation approach, investigators identify time zero for each exposed cluster at policy adoption or implementation and then create a corresponding comparison group from clusters that have not yet been treated (21,22). By contrast, for the proposed SW-CRT emulation approach, a single calendar time representing the beginning of the study is used for all clusters. Rather than aligning time zero with the implementation time, the intervention status for each cluster defines whether each period is labeled exposed or unexposed over time. Design and analysis will therefore accommodate this time-varying exposure and account for secular trends as in SW-CRT analyses.

*Component 5. Outcomes.* For both the target trial and emulation, the outcomes are defined, as well as how they are measured and aggregated. For SW-CRT emulation, understanding the individuals on whom outcomes are measured in each cluster-period is important to distinguish closed-cohort from cross-sectional data (15).

*Component 6. Causal Contrasts.* The causal contrasts or estimands of interest are clearly specified for the target trial, including choosing cluster- or individual-level estimands, how effects are averaged across individuals and clusters, the scale of the contrast (e.g., multiplicative or additive), and the handling of any missing data or nonadherence at the cluster- or individual-level (10,32). Note that these align with design and estimand reporting guidance for CRTs and SW-CRTs (30–32). For the observational analog, similar considerations are needed, as well as whether any additional conditioning is used in the estimand (e.g., to adjust for necessary covariates), whether the estimand is an average treatment effect in the entire study population, average treatment effect among the treated, or some other measure, and how changes to the policy after implementation are treated.

For CRTs, informative cluster size can play an important role in selecting and interpreting estimands, especially the distinction between cluster-specific and population-average estimands (11,36,37). In addition, for SW-CRTs, how any time-varying or dynamic treatment effects are modeled, aggregated, and weighted are key considerations and relevant state-of-the-art techniques should be considered (6,29,38–40). The SW-CRT emulation clarifies how assumptions on effect heterogeneity affect the ability to adjust for period effects efficiently (29). These decisions should inform the observational analog.

*Component 7. Identifying Assumptions.* Assumptions for identification of the trial's causal estimand may include: constant treatment effects across time and/or units, exchangeability of units in time trends, perfect adherence, no spillover effects, no anticipation effects, random sampling of individuals within clusters, and others (15). In addition to these, the observational analog will likely require additional exchangeability assumptions that address the non-random timing of policy adoption: parallel trends or conditional parallel trends for DID-based analyses, linear interpolation and stable weights for synthetic control, or conditional exchangeability for confounder-adjusted analyses (2,4,5,9).

*Component 8. Data Analysis Plan.* Based on the causal contrasts of interest and identifying assumptions, the target trial analysis method is pre-specified. This includes modeling approaches to account for time trends, correlated outcomes, missing data, and excluded cluster-periods (41). For the observational analog, the data analysis plan will also include approaches to approximate exchangeability among clusters with different treatment patterns; for example, DID with covariate adjustment to achieve conditional parallel trends, matching or weighting of clusters, synthetic controls, or robust combinations of these approaches (4,6,7). Study investigators should align their reporting to SW-CRT guidance when possible (31,32,42).

**Examples: Vaccine Incentives**

To demonstrate the use of this framework and relevant SW-CRT literature for staggered adoption settings, we describe two examples assessing the effect of state-level policies to encourage COVID-19 vaccination in 2021.

*Example 1: Vaccine Lottery*

The first policy, a vaccine lottery, provided an entry to win a large financial prize to those who received their first dose of COVID-19 vaccination prior to or within a set period after the announcement of the lottery. Twenty states implemented such policies with varying prizes. Analyses of these policies using DID and synthetic control approaches have so far provided mixed evidence, although there does seem to be evidence of increased first-dose vaccinations in the short run after announcement (43–45).

Here, we demonstrate the use of this TTE framework to design a study to assess the short-term effect of state-level vaccine lotteries on first-dose vaccination rates. We generally use the policy definitions and outcome data availability described by Fuller et al. (43). Table 1 summarizes one possible set of TTE components applied to this analysis. This setting has important time trends in the outcome (new first-dose vaccinations) and likely heterogeneity in effects by state, calendar time (as baseline vaccination rates and local epidemic conditions change), and time since implementation. The SW-CRT emulation approach thus allows us to use information efficiently across states and identify relevant estimands.

| Component | Hypothetical Target Trial | Observational Study Analog/Emulation |
|---|---|---|
| 1. Eligibility Criteria | U.S. states with first-dose vaccination rates among adults less than 70% as of May 8, 2021, excluding those implementing broad-based vaccine mandates (randomization and analysis units). | U.S. states with first-dose vaccination rates among adults less than 70% as of May 8, 2021, excluding those implementing broad-based vaccine mandates (policy and measurement units). |
| 2. Treatment Strategies | Intervention Condition: Policy offering one $1,000,000 prize to a randomly chosen individual who received their first COVID-19 vaccine dose prior to six weeks after lottery announcement, announced between May and July 2021. Control Condition: No financial incentive (lottery or guaranteed) policy for COVID-19 vaccination announced yet. | Exposed Condition: Policy offering one or more prizes of at least $100,000 to a randomly chosen individual who received their first COVID-19 vaccine dose prior to some set period after lottery announcement, announced between May and July 2021. Control Condition: No financial incentive (lottery or guaranteed) policy for COVID-19 vaccination announced yet. |
| 3. Assignment Procedures | Stepped-wedge (randomized staggered roll-out) randomization with multiple randomized timing sequences and a baseline group. Randomization would have to be blinded until the time period of announcement. | Non-randomized staggered adoption times by standard political means, with various announcement dates or no policy announced during study period. Exchangeability of adoption times through covariate adjustment, matching/weighting, or synthetic controls, with additional assumptions. |
| 4. Follow-Up | Duration: CDC MMWR Weeks 15–30 (ending April 17–July 31, 2021) (46). Time Zero: the study start period (first outcome measurement collected). Excluded Period(s): the week of announcement (implementation or washout period). | Duration: CDC MMWR Weeks 15–30 (ending April 17–July 31, 2021). Time Zero: the study start period (first outcome measurement used). Excluded Period(s): the week of announcement. |
| 5. Outcomes | State-level adult (18+) first-dose vaccinations per week, as a percentage of the total adult population. | State-level adult (18+) first-dose vaccinations per week, as a percentage of the total adult population. |
| 6. Causal Contrasts | The intention-to-treat difference in weekly percentage receiving a first-dose vaccination, averaged over three weeks post-implementation (average treatment effect). Includes all randomized states. | The average treatment effect *on the treated*, averaged over timing groups and three weeks post-implementation, in percentage receiving a first-dose vaccination. The treated includes all states that announced a lottery by study end. |
| 7. Identifying Assumptions | Allows time-on-treatment effect heterogeneity; assumes no calendar time heterogeneity and does not model unit effect heterogeneity explicitly. Assumes one week of implementation period and no anticipation or spillover effects. | Allows time-on-treatment effect heterogeneity and averages over calendar time effect heterogeneity; assumes unit effect homogeneity. Assumes one excluded week and no anticipation or spillover effects. Assumes that either the outcome regression or inverse probability of treatment weighting model sufficiently guarantees parallel trends. |
| 8. Data Analysis Plan | Linear mixed effects model with state-level random effects, time fixed effects, and state- | Callaway and Sant'Anna doubly-robust DID for staggered adoption with cluster-level data |

| | time random effects with treatment by time-on-treatment interaction, averaging over three post-implementation periods (39,41). Inference by permutation test and asymptotic model-based 95% CI. | (6), with week 18 population percentage of already-vaccinated or vaccine-refusing as covariate and the not-yet-treated as controls. Aggregated by average across three post-implementation periods and timing groups. Inference by placebo test and cluster-bootstrap 95% CI. |
|---|---|---|

*Table 1*. An example of the target trial emulation framework applied to the analysis of U.S. state-level vaccine lotteries by emulating a stepped-wedge cluster randomized trial.

Several components have alternative design and analysis options. For components 1 and 6, the eligible states and time periods could be expanded (e.g., to longer term effects) or restricted (e.g., to specific states; see Figure 1). This affects variance and generalizability in both the randomized and observational settings and affects exchangeability in the observational setting. Parallel trends, for example, may be more reasonable on a restricted set of similar states or on a shorter time scale (7,47). For component 2, a stricter definition could be used for the exposed condition in the observational study; this could improve consistency but reduce the sample size. For component 5, full vaccination could be used as the outcome instead of first-dose; this may be more public health-relevant but occurs at a greater remove from the policy, risking biases. For components 7 and 8, different assumptions of effect heterogeneity or conditional parallel trends could be specified, leading to different estimators and analysis plans.

A) Schematic of design using all Midwest region states

| State | 15 | 16 | 17 | 18 | 19 | 20 | 21 | 22 | 23 | 24 | 25 | 26 | 27 | 28 | 29 | 30 |
|---|---|---|---|---|---|---|---|---|---|---|---|---|---|---|---|---|
| OH | | | | | May 12, 2021 | | | | | | | | | | | |
| IL | | | | | | | | | | | June 17, 2021 | | | | | |
| MI | | | | | | | | | | | | | June 30, 2021 | | | |
| MO | | | | | | | | | | | | | | | July 21, 2021 | |
| IA | | | | | | | | | | | | | | | | |
| IN | | | | | | | | | | | | | | | | |
| KS | | | | | | | | | | | | | | | | |
| MN | | | | | | | | | | | | | | | | |
| ND | | | | | | | | | | | | | | | | |
| NE | | | | | | | | | | | | | | | | |
| SD | | | | | | | | | | | | | | | | |
| WI | | | | | | | | | | | | | | | | |

B) Schematic of design using a matched control state for each intervention state

| State | 15 | 16 | 17 | 18 | 19 | 20 | 21 | 22 | 23 | 24 | 25 | 26 | 27 | 28 | 29 | 30 |
|---|---|---|---|---|---|---|---|---|---|---|---|---|---|---|---|---|
| OH | | | | | May 12, 2021 | | | | | | | | | | | |
| ND | | | | | | | | | | | | | | | | |
| IL | | | | | | | | | | | June 17, 2021 | | | | | |
| MN | | | | | | | | | | | | | | | | |
| MI | | | | | | | | | | | | | June 30, 2021 | | | |
| KS | | | | | | | | | | | | | | | | |
| MO | | | | | | | | | | | | | | | July 21, 2021 | |
| IN | | | | | | | | | | | | | | | | |

C) Schematic of design using only states that adopt the policy ("timing-only" comparisons)

| State | 15 | 16 | 17 | 18 | 19 | 20 | 21 | 22 | 23 | 24 | 25 | 26 | 27 | 28 | 29 | 30 |
|---|---|---|---|---|---|---|---|---|---|---|---|---|---|---|---|---|
| OH | | | | | May 12, 2021 | | | | | | | | | | | |
| IL | | | | | | | | | | | June 17, 2021 | | | | | |
| MI | | | | | | | | | | | | | June 30, 2021 | | | |
| MO | | | | | | | | | | | | | | | July 21, 2021 | |

*Figure 1.* Design schema for alternative unit inclusion criteria (Component 1). *MMWR* weeks are as defined by the U.S. Centers for Disease Control and Prevention. Shaded boxes indicate post-adoption time periods; dates indicate lottery announcement dates for the relevant states.

The emulation can allow us to quantify potential trade-offs as well. Using pre-intervention data (43), we can estimate baseline rates of vaccine uptake, a cluster-specific random effect, and period effects for this setting under a mixed effects model (48). We can also estimate feasible effect sizes based on prior studies. We hypothesize an effect size of 0.33% per week based on scaling down the effect estimated for a more targeted policy (49). Using the *swdpwr* package in R (50), a SW-CRT analysis of the Midwest region with the observed sequence of crossovers would have 78% power to detect this effect. Restricting to only four better-matched control states would give 61% power, while excluding all non-adoption states would give 38% power (see Figure 1 schematics and see Appendix 1 for details).

While these values are not precise for analysis methods used for non-randomized studies as inference approaches generally differ, they inform feasibility and provide a comparison for the variance trade-off from restricting the set of states included in the analysis. The risk of bias when using never-treated units or a larger pool of control units has been noted for staggered adoption DID and synthetic controls (7,8,47,51). These calculations help quantify the reduction in power that arises from restricting the comparison set to potentially reduce bias. In this example, the investigator may feel that the matched comparison set is worth the modest reduction in power estimated from the calculations. If the third option (only including intervention states) is necessary to ensure exchangeability, gathering data for more time points or identifying a less variable outcome may be needed for adequate power.

*Example 2: State Employee Mandate*

Another set of policies adopted in some U.S. states in 2021 were vaccine mandates for state employees. These policies exhibited heterogeneity regarding who was subject to them, at what time point, strictness of enforcement, and opt-out provisions (52). Nonetheless, DID analyses to assess the effects of these policies on overall COVID-19 vaccine uptake in that season, as well as vaccination rates in the next annual season, found null and negative effects, respectively (53). This contradicted existing research showing positive effects of various employer mandates on vaccination (54), and the methodology has been questioned (55).

Considering the TTE framework, a hypothetical SW-CRT for this analysis would randomize an order for policy implementation across states, with perhaps a group that never implements the policy (14). Table 2 provides three examples of limitations of this study identified through TTE.

| Component | Challenge in Hypothetical Stepped-Wedge Cluster Randomized Trial | Additional Challenge in Observational Study Emulation |
|---|---|---|
| 2. Treatment Strategies | Ensuring consistent implementation of vaccine mandate timing and enforcement across clusters | High policy variation, with uneven publicity, timing, enforcement, and rescinding of policies. |
| 4. Follow-Up | Next-season analysis has very long duration, creating highly variable outcomes and potential mediators and effect modifiers that differ across clusters. | Lengthy duration allows intermediate policy changes and confounding effect modifiers due to non-randomized adoption (55). Timing-only comparisons are of limited use because follow-up duration is much longer than timing variation (5). |
| 1. Eligibility Criteria; 5. Outcomes | Aggregate state-level vaccination rates for adults include many individuals not directly affected by the policy, leading to high variation of the outcome compared to possible effect size. | Additional correlated policies or confounding factors may influence the outcomes of this larger population. |

*Table 2*. Selected challenges of an observational study of U.S. state employee vaccine mandate policies identified by emulating a stepped-wedge cluster randomized trial.

While TTE is not the only way to identify these (and other) challenges for this observational study, it helpfully frames them. This can highlight limitations of the proposed study and engender appropriate skepticism toward the results. The TTE here suggests the data are not well-suited to answer the question. This, along with specific critiques of the analysis method used (55), can inform what analyses are conducted and how findings are interpreted and used by policymakers (1).

**Discussion**

CRTs and SW-CRTs are useful designs to understand the effect of group-level interventions, including health policies. Using them as target trials for observational studies helps clarify the opportunities and limitations of such studies, can identify infeasible settings, and provides transparency to strengthen the rigor and interpretation of analyses.

Explicitly emulating SW-CRTs—rather than aligning time zero and emulating parallel-arm CRTs (21,22)—provides clarity on the time-varying assignment mechanisms and intervention effects and may point to more powerful analysis approaches. First of all, it makes clear how much information is available to the investigator overall, including for estimating the confounding period and unit effects (35,56,57). Secondly, it can provide possible methods for approximate power calculations or comparisons (41,48,50). Thirdly, it can point to assumptions or models for treatment effect heterogeneity that may allow for improved estimation, power, and interpretation of desired effects (29,38,39,56,58). It can also highlight other potential biases considered in SW-CRTs, including implementation heterogeneity, spillover, and anticipation (15). Finally, it encourages consideration of key limitations of CRTs and SW-CRTs more generally; frameworks for reporting these can improve reporting of observational studies as well (15,59).

TTE can inform the trade-offs resulting from various design decisions through relevant CRT literature. The trade-off among bias, variance, and generalizability can be especially pronounced for CRTs and SW-CRTs with few units of randomization (59,60). A simple schematic of the treated and control periods, common for SW-CRTs (e.g., Figure 1), improves transparency and can be supplemented with consideration of the information content of cells (35,56). In longitudinal CRT designs, extensive work has analyzed the variance and precision consequences of, for example, different cohort recruitment approaches, increased implementation periods and design types and randomization schedules (12,34,35). In some cases, preliminary power or sample size calculations using SW-CRT methods may provide guidance on whether the quasi-experimental study will have reasonable power, helping prevent waste of research resources or misleading null results (41,48,50,61,62). This is especially useful given that power calculations, unlike identifiability conditions, have not been a major focus of the recent observational staggered adoption literature (4,7). These calculations help quantify design considerations, such as the risk of bias with a wider pool of units versus the improved precision and power that such inclusion may bring. Further research on appropriate use of these calculations is warranted.

Treatment effect heterogeneity—including dynamic effects, effects that vary by calendar time, and group- and individual-level heterogeneity—has been a theme of recent literature in both fields (5–7,29,38–40,58). This provides an opportunity for these areas to learn from one another. For quasi-experiments, the focus has largely been on bias that can arise from dynamic effects, and the lack of clear interpretation of the weighted average estimator, with various proposed solutions (5–7). For SW-CRTs, there has been research on these weights, as well as development of models that capture these dynamic effects or that are robust to them and methods to detect the heterogeneity itself with adequate power (29,38,39,57,58,63–67). The bias-variance-generalizability (and interpretability) trade-offs can best be assessed by considering all of these features together in specific contexts. Investigators may thus identify settings, both randomized and observational, where the dynamic treatment effects are small enough to be negligible or settings where they are so severe as to call into question the validity of a staggered adoption approach at all.

Other threats to validity and generalizability, such as lack of consistency (33), are considered in both areas as well, and research would benefit from considering the insights in both the quasi-experiments and CRT literature. For example, the possibility of improved implementation at later time points—or, more generally, heterogeneity of implementation quality—is a common consideration in pragmatic SW-CRTs (68). This may be a similar consideration for evaluating states that adopt policies after viewing them in other states (e.g., states adopting vaccine lotteries after Ohio did) (43). Spillover and bypass effects are also key issues that can arise in both randomized and observational settings (e.g., individuals near exposed states being influenced by those policies) (44). This difficulty of identifying units that are similar but do not risk spillover is common to both CRTs and quasi-experiments (10,47,69). While the TTE framework does not solve these challenges, emulating a trial encourages transparent consideration of them.

The TTE framework does not, in itself, provide information on how successful the confounding adjustment or emulation will be. It may, however, help assess the validity of required assumptions; for example, horizontal-focused estimators (63,70) or timing-only estimators that use only not-yet-treated clusters as controls (5,6). Untestable exchangeability assumptions, however, remain a fundamental threat to observational studies and should not be ignored (18). TTE can also reproduce biases that exist in randomized trials or crowd out alternative approaches to causal inference (71,72). Moreover, it should not be used to minimize the need for triangulation of different forms of evidence, which may be especially pronounced for policy effects where there are many important causal questions (1,71,73).

Emulating SW-CRTs can help communicate these challenges even more explicitly and allow for careful consideration of policy adoption timing and time-varying effects. Extending these ideas to other longitudinal CRT designs, or related designs such as staircase CRTs (12,74), can provide additional approaches useful for particular settings. CRTs and SW-CRTs are often imperfect and designed to accommodate practical considerations and challenges (10,13,15,59); quasi-experimental analyses of policies will be as well. Instead of providing a single "gold-standard" answer, this framework should be approached with consideration of the scientific goals, feasibility, and interpretability and usefulness of results. By encouraging transparent choices around estimands, assumptions, and methods, it encourages stronger reporting and a clearer assessment of the value and limitation of the evidence generated.

**Acknowledgements**

H.L. was supported by National Institutes of Health (NIH)/National Institute on Drug Abuse (NIDA) under grant number (R00DA057487). G.G. was supported by National Institutes of Health (NIH)/National Institute on Drug Abuse (NIDA) under grant number (R01DA060716). F.L. was supported by the Patient-Centered Outcomes Research Institute® (PCORI® Award ME-2022C2-27676) and the United States National Institutes of Health (NIH), National Heart, Lung, and Blood Institute (NHLBI, grant number 1R01HL178513). The statements presented in this article are solely the responsibility of the authors and do not necessarily represent the views of PCORI®, its Board of Governors, Methodology Committee, or NIH. The authors declare that there are no conflicts of interest relevant to this work.

**Appendix 1**

We generate example power calculations for three possible study designs based on the target trial emulation of the vaccine lottery policies. A Github repository with the relevant data and code is available at https://github.com/leekshaffer/SWT_TTE_Appx.

The data are collated from that provided by Fuller et al. (43), using CDC MMWR Weeks to index the time periods (46). See the Harvard Dataverse repository (https://doi.org/10.7910/DVN/K1XX02) for the original data.

The three designs are as follows:

1. Use all states in the CDC-defined Midwest region (12 total states), with all observations from CDC MMWR Weeks 15–30 of 2021.
2. Use the four intervention states in the CDC-defined Midwest region (Ohio, Illinois, Michigan, Missouri) and a matched-comparison state for each, with all observations from CDC MMWR Weeks 15–30 of 2021.
3. Use only the four intervention states in the CDC-defined Midwest region, with all observations from CDC MMWR Weeks 15–30 of 2021.

The schematics for these designs are shown in Figure 1 of the main text. Note that these are only examples of designs that could be considered to illustrate how stepped-wedge trial power calculations can be used in the target trial emulation framework to inform the design of quasi-experimental studies.

*Identifying matched comparison states for design 2*

To identify matched comparison states for design 2, we consider three covariates collected through administrative and survey data of each state in Week 18, the final week prior to the earliest vaccine lottery adoption (in Ohio) (43). These covariates are:

- The already vaccinated percentage (W18_First_18Pop_Pct) calculated as the percentage of the over-18 population already vaccinated (with at least one dose).
- The excluded percentage (W18_Excl_Perc) calculated as the already vaccinated percentage (W18_First_18Pop_Pct) plus the percentage of the over-18 population who responded that they definitely will not get vaccinated among those not yet vaccinated.

- The persuadable percentage (W18_Persu_Perc) calculated as the percentage of the over-18 population who has not yet been vaccinated and did not respond that they definitely or probably will not get vaccinated.

For each intervention state, the Euclidean (sum of squares) distance to each control state is calculated across these covariates. The control state with the minimum distance is used as the matched comparison state. These matches are given in Table S1.

| Intervention State | Matched Comparison State |
|---|---|
| Ohio (OH) | North Dakota (ND) |
| Illinois (IL) | Minnesota (MN) |
| Michigan (MI) | Kansas (KS) |
| Missouri (MO) | Indiana (IN) |

*Table S1. Intervention and matched comparison states for design 2*

*Estimating parameters from control data*

We estimate several parameters needed for the power calculations from the data. For the expected outcomes, we estimate the mean outcome under control at the start of the study (Week 15), at the end of the study (Week 30), and under intervention at the end of the study (Week 30). The outcome considered is the percentage of the population receiving their first dose of the COVID-19 vaccine in the week.

The mean responses under control are estimated by fitting an interrupted linear model for this outcome by week on the control observations. We model a linear trend for time, with a changepoint after week 20 (determined based on visual inspection). From this, we estimate a mean outcome of 3.39 percentage points in Week 15 and 0.48 percentage points in Week 30.

An individually randomized trial of direct financial incentives found a 4 percentage point increase in uptake over 30 days from a baseline of around 70% uptake (49). We estimate power for an effect size 1/3 of that (0.33% per week) for a broadly-targeted lottery.

To estimate variance parameters, mixed effects models are fit on the data from the control periods. These are fit with fixed effects of week (categorical terms) and random intercepts for states (48). The marginal variance and estimated intracluster correlation coefficient (ICC) are obtained. This is done for design 1, using all twelve states, and design 2, using only the four intervention states and their four matched comparison states. As design 3 has only four states, with limited control observations in each, it does not give stable ICC estimates. For this design, we use those for design 2.

The estimated design parameters used are given in Table S2.

| Parameter | swdpower() argument name (50) | Design 1 | Design 2 | Design 3 |
|---|---|---|---|---|
| Sample Size (Clusters) | | 12 | 8 | 4 |
| Sample Size (Observations) | | 192 | 128 | 64 |
| Expected Control Outcome at Baseline (Week 15) | meanresponse_start | 3.39 | | |
| Expected Control Outcome at End (Week 30) | meanresponse_end0 | 0.48 | | |
| Expected Intervention Outcome at End (Week 30) | meanresponse_end1 | 0.81 | | |
| Marginal Outcome Variance | sigma2 | 0.26 | 0.35 | |
| Significance Level | typeIerror | 0.05 | | |
| Intracluster Correlation Coefficient | alpha1 | 0.39 | 0.42 | |

*Table S2. Estimated design parameters for considered designs*

*Design schematics*

Data sets representing the design schematics (i.e., the included cluster-periods and their intervention status) are created. These are shown in Figure 1 of the main text. Note that the ordering of clusters does not affect power calculations.

*Power calculations*

We use the *swdpwr* package in R (50) to estimate the power these designs would have, using the estimated parameters described above, if they were conducted as stepped-wedge cluster-randomized trials. Note, as discussed in the main text, that these are not necessarily accurate for the analysis methods that would be used in the observational analysis, so they are used to guide feasibility considerations and understand trade-offs. The code for these calculations is available at https://github.com/leekshaffer/SWT_TTE_Appx.

The results are given in Table S3 (and discussed in the main text).

| Design | Sample Size | Estimated Power |
|---|---|---|
| Design 1 | 12 states: 4 intervention, 8 control | 78% |
| Design 2 | 8 states: 4 intervention, 4 matched controls | 61% |
| Design 3 | 4 states: intervention only | 38% |

*Table S3. Estimated power for considered designs, using* swdpwr *and described parameters*